\documentclass{emulateapj} 
\newcommand{\msun}{\mbox{M$_{\sun}$}}

\newcommand{\bl}[1]{\mbox{\boldmath$ #1 $}}

\slugcomment{Submitted to Astrophysical Journal}
\shorttitle{Disk masses around young stellar objects}
\shortauthors{Vorobyov} 

\begin{document}

\title{Disk masses in the embedded and T Tauri phases of stellar evolution}
\author{E. I. Vorobyov\altaffilmark{1,}\altaffilmark{2}}
\altaffiltext{1}{Institute for Computational Astrophysics, Saint Mary's University,
Halifax, B3H 3C3, Canada; vorobyov@ap.smu.ca.} 
\altaffiltext{2}{Institute of Physics, South Federal University, Stachki 194, Rostov-on-Don, 
344090, Russia.} 

%\maketitle

\begin{abstract}
Motivated by a growing concern that masses of circumstellar disks may
have been systematically underestimated by conventional observational methods, 
we present a numerical hydrodynamics study of 
time-averaged disk masses ($\langle M_{\rm d} \rangle$) around low-mass
Class~0, Class~I, and Class~II objects. Mean disk masses ($\overline{M}_{\rm d}$)
are then calculated by weighting the time-averaged disk masses according to the 
corresponding stellar masses using a power-law weight function with a slope 
typical for the Kroupa initial mass function of stars.
Two distinct types of disks are considered: self-gravitating disks,
in which mass and angular momentum are redistributed exclusively by gravitational torques,
and viscous disks, in which both the gravitational and viscous torques are at work.
We find that self-gravitating disks have mean masses that are slowly increasing along the sequence
of stellar evolution phases. More specifically, Class~0/I/II self-gravitating 
disks have mean masses $\overline{M}_{\rm d}=0.09$, $0.10$, and $0.12~M_\odot$, respectively.
Viscous disks have similar mean masses ($\overline{M}_{\rm d}=0.10-0.11~M_\odot$) 
in the Class 0/I phases but almost a factor of 2 lower mean mass in the Class II phase 
($\overline{M}_{\rm d,CII}=0.06~M_\odot$). In each evolution phase, time-averaged disk masses
show a large scatter around the mean value.
Our obtained mean disk masses are {\it larger} than
those recently derived by Andrews \& Williams and Brown et al., regardless of the 
physical mechanisms of mass transport in the disk. 
The difference is especially large for Class II disks, 
for which we find $\overline{M}_{\rm d,CII}=0.06-0.12~M_\odot$ but Andrews and Williams 
report median masses of order $3\times 10^{-3}~M_\odot$. 
When Class 0/I/II systems are considered altogether, a least-squares best fit yields 
the following relation between the time-averaged disk and
stellar masses, $\langle M_{\rm d} \rangle = \left( 0.2\pm 0.05 \right) 
\langle M_\ast \rangle^{1.3\pm 0.15}$.  
The dependence of $\langle M_{\rm d} \rangle$ on $\langle M_\ast \rangle$ 
becomes progressively steeper along the sequence of stellar evolution phases,
with exponents $0.7\pm 0.2$, $1.3\pm 0.15$, and $2.2\pm 0.2$ for Class~0, 
Class~I, and Class~II systems, respectively.
\end{abstract}


\section{Introduction}
It has now become evident that disks of gas and dust are present from the earliest 
phases of stellar evolution (Class~0 and Class~I) and last for at least several million years
into the late Class~II phase. Disks are observed or inferred around most T Tauri stars and 
even around brown dwarfs. 
The evidence for disks around Class~0 and Class~I sources is more indirect.
In this early phase of stellar evolution, the protostar/disk system is deeply 
embedded in an envelope -- a remnant of the cloud core from which the protostar is forming. 
Nevertheless, recent observations by \cite{Andrews} suggest that Class I disks
have a larger median mass than that of Class II disks.

In spite of a considerable progress in the detection of disks around young stellar objects 
(YSOs), an accurate determination of disk masses is still challenging.
It is difficult to directly determine disk masses from the spectral lines 
of molecular species because the brightest, easily detectable lines 
(i.e., the rotational transitions of CO) are optically thick and likely to be severely 
depleted. Therefore, disk masses are usually inferred from analyzing the spectral 
energy distribution of YSOs from the mid-infrared through submillimeter bands 
\citep[e.g.][]{Andrews,Brown}. Such measurements of disk masses suffer from large 
uncertainties in the normalization of dust opacities and gas-to-dust ratios,
which led \citet{Hartmann} to conclude that T Tauri disk masses have been 
systematically underestimated by conventional analyses.

Another complication arises from poorly known physical processes in the disk.
A usual assumption of optically thin circumstellar disks may significantly 
underestimate disk masses, particularly for objects with larger flux densities.
However, a self-consistent treatment of a non-negligible optical depth requires 
a knowledge of the radial gas surface density profile in the disk \citep{Andrews}, 
which may depend significantly on the dominant physical mechanism of mass and angular 
momentum redistribution in the disk \citep{VB4}.

Given large uncertainties in the measurements of disk masses, numerical simulations
of self-consistent formation and evolution of circumstellar disks can provide  
valuable information on disk masses in the early embedded and late phases 
of YSO evolution. It has been shown in the past that the saturation of spiral
gravitational instabilities at a finite amplitude in a self-gravitating, Toomre-unstable disk 
allows for the steady transport of mass and momentum, which eventually limits
disk masses \citep[e.g.][]{Adams89,Shu90,Laughlin97,Laughlin98}.
In this paper, we perform numerical simulations of
the long-term evolution of self-consistently formed circumstellar disks
around low-mass stars ($0.2~M_\odot \la M_\ast\la 2.0~M_\odot$).
We consider both the self-gravitating disks, in which radial transport of mass and
angular momentum is done exclusively via gravitational torques, and viscous disks, which feature 
gravitational torques as well as viscous ones. 
We seek to determine numerically the disk masses in the Class 0, Class I, and Class II phases
of stellar evolution.

The paper is organized as follows. 
The numerical methods and initial parameters of cloud cores are 
given in \S~\ref{methods}. Our obtained masses for self-gravitating and viscous disks are
presented in \S~\ref{SG} and \S~\ref{VD}. We compare our numerical results with observations in 
\S~\ref{discuss}. The model and numerical caveats are discussed in \S~\ref{caveats}. 
The main results are summarized in \S~\ref{summary}.

\section{Description of numerical methods}
\label{methods}
We use the thin-disk approximation to compute the evolution of rotating, 
gravitationally bound cloud cores. This allows efficient calculation of 
the long-term evolution of a large number of models.
We start our numerical integration in the pre-stellar phase, which is characterized 
by a collapsing {\it starless} cloud core, and continue into the main accretion phase,
which sees the formation of a central star and circumstellar disk.
We cover all major phases of the evolution of a YSO, starting from 
its formation and ending with the T Tauri phase. The integration ends when the age of 
the central star is about three million years. In some models we extend the integration
up to 5~Myr. We emphasize that circumstellar disks
are formed self-consistently in our numerical simulations, 
rather than being introduced as an initial parameter of the model.

Once the disk is formed, its mass is determined by an interplay between the efficiency 
of the mass and angular momentum transport in the disk\footnote{In fact, disks may also transport
angular momentum to the external environment due to magnetic braking. This effect will be 
considered in a follow-up paper.} and the infall rate of matter from the surrounding envelope onto the
disk. At the time of disk formation, the infall rates take values between 
$1.2\times 10^{-6}~M_\odot$~yr$^{-1}$ and  $7.0\times 10^{-6}~M_\odot$~yr$^{-1}$  (measured at $600$~AU)
for the least and most massive cloud cores, respectively, but they show a fast decline with time.
These values (and strong time variation) are consistent with the infall rates 
derived by \citet{Klessen01} using 
numerical models that follow molecular cloud evolution from turbulent fragmentation toward the 
formation of stellar clusters. We note that once the disk is formed, 
the infall rate of matter from the envelope onto the disk is not necessarily the same 
as the mass accretion rate from the disk onto the protostar. While the former shows a fast decline
with time, the latter is usually characterized by a much slower decline and has a strong dependence
on the stellar mass \citep{VB3,VB5}.

We use two numerical approaches: a basic approach that accounts for the radial transport due to gravitational
toques and a viscous approach that accounts for the radial transport due to both the gravitational
torques and viscosity.
Gravitational torques are known to efficiently redistribute mass and 
angular momentum in circumstellar disks \citep[e.g.][]{Lodato04,Lodato05}. 
They were shown to play an important role in driving the FU-Ori-like bursts
in the early embedded phase of disk evolution \citep{VB1,VB2}. In the late disk evolution, 
negative gravitational torques associated with low-amplitude azimuthal density perturbations
in the disk can drive mass accretion rates that are consistent with those measured in 
the intermediate and upper-mass T Tauri stars \citep{VB3,VB5}.

\subsection{Basic numerical approach}
\label{basic}

In the basic numerical approach, the collapse of a cloud core and subsequent evolution 
of a star/disk system 
is carried out by solving the basic equations of mass and momentum transport 
in the thin-disk approximation \citep[see e.g.][]{VB2}
%are derived by integrating the ideal hydrodynamics equations in the z-direction,
%from $z = Z(r, \phi, t)$ to $z = -Z(r, \phi, t)$, and using Leibniz's rule
%for differentiation of integrals and the fundamental theorem of
%calculus \citep[see e.g.][]{VB2}
\begin{eqnarray}
\label{cont}
 \frac{{\partial \Sigma }}{{\partial t}} & = & - \nabla _p  \cdot \left( \Sigma \bl{v}_p 
\right), \\ 
\label{mom}
 \Sigma \frac{d \bl{v}_p }{d t}  & = &  - \nabla _p {\cal P}  + \Sigma \, \bl{g}_p \, ,
\end{eqnarray}
where $\Sigma$ is the mass surface density, ${\cal P}=\int^{Z}_{-Z} P dz$ is the vertically integrated
form of the gas pressure $P$, $Z$ is the radially and azimuthally varying vertical scale height,
 $\bl{v}_p=v_r \hat{\bl r}+ v_\phi \hat{\phi}$ is the velocity in the
disk plane, $\bl{g}_p=g_r \hat{\bl r} +g_\phi \hat{\phi}$ is the gravitational acceleration 
in the disk plane, and $\nabla_p=\hat{\bl r} \partial / \partial r + \hat{\phi} r^{-1} 
\partial / \partial \phi $ is the gradient along the planar coordinates of the disk. 
The gravitational acceleration $\bl{g}_p$ is found by solving for the Poisson integral 
\citep[see][]{VB2}. The fact that we account for the disk self-gravity means
that gravitational torques arise {\it self-consistently} in our numerical simulations and
not imitated by some means of $\alpha$-viscosity. 

Taking into account the complexity of gas thermodynamics in circumstellar disks
(see \S~\ref{caveats}), we have adopted a barotropic 
equation of state that closes equations~(\ref{cont}) and (\ref{mom}) and makes 
a transition from isothermal to adiabatic evolution at $\Sigma = \Sigma_{\rm cr} = 
36.2$~g~cm$^{-2}$
\begin{equation}
{\cal P}=c_s^2 \Sigma +c_s^2 \Sigma_{\rm cr} \left( \Sigma \over \Sigma_{\rm cr} \right)^{\gamma},
\label{barotropic}
\end{equation}
where $c_s$ is the isothermal sound speed, the value of which is set equal to that 
of the initial cloud core, and $\gamma=1.4$. 
Equation~(\ref{barotropic}), though neglecting detailed
cooling and heating processes, was shown to reproduce to a first approximation 
the radial temperature gradients in the disk \citep{VB3} and the density-temperature relation 
for collapsing cloud cores derived by \citet{Masunaga} using a detailed radiation hydrodynamics 
simulation \citep{VB2}.

The vertical scale height $Z(r,\phi,t)$ is determined assuming the local hydrostatic 
equilibrium in the gravitational field of a disk and central star \citep{VB4}.
The relevant formulas are given in the Appendix.

\subsection{Viscous numerical approach}
\label{viscmodel}
Viscosity is another important mechanism of angular momentum 
and mass redistribution in astrophysical disks. Most analytical and numerical studies 
of viscous evolution of thin circumstellar disks have employed 
the standard axisymmteric model of \citet{Lyndenbell}, in which 
the surface density of a Keplerian disk evolves with time according to the following diffusion equation
\begin{equation}
{\partial \Sigma \over \partial t} = {3\over r} {\partial \over \partial r} 
\left[ r^{1/2} {\partial \over \partial r} ( \nu \Sigma r^{1/2}) \right],
\end{equation}
where $\nu$ is the kinematic viscosity. 

In the present paper, we take a more fundamental approach and describe the effect of (yet unspecified)
viscosity in terms of the classic viscous stress tensor 
\begin{equation}
\mathbf{\Pi}=2 \mu \left( \nabla v - {1 \over 3} (\nabla \cdot v) \mathbf{e} \right),
\end{equation}
where $\nabla v$ is a symmetrized velocity gradient tensor, $\mathbf{e}$ is the unit tensor, and
$\mu$ is the dynamical viscosity.
This approach allows for a 
self-consistent treatment of both, self-gravity and viscosity, within the same numerical formalism.
The resulting mass and momentum transport equations in the viscous numerical approach are
\begin{eqnarray}
\label{visc1}
 \frac{{\partial \Sigma }}{{\partial t}} & = & - \nabla _p  \cdot \left( \Sigma \bl{v}_p 
\right), \\ 
\label{visc2}
 \Sigma \frac{d \bl{v}_p }{d t}  & = &  - \nabla _p {\cal P}  + \Sigma \, \bl{g}_p
 + \left( \nabla \cdot \mathbf{\Pi}\right)_p \, ,
\end{eqnarray}
where $\nabla \cdot \mathbf{\Pi}$ is the divergence of the rank-two viscous stress tensor 
$\mathbf{\Pi}$. The relevant components of $\left( \nabla \cdot \mathbf{\Pi} \right)_p$ 
are given in the Appendix.
Equations~(\ref{visc1}) and (\ref{visc2}) are closed with the barotropic equation of state~(\ref{barotropic}).
We note that the viscous approach accounts self-consistently for both the gravitational and 
viscous torques which may arise during numerical simulations.

It is evident that the practical application of equation~(\ref{visc2}) requires a knowledge
of the dynamical viscosity $\mu$ of the disk.
Unfortunately, our understanding of viscous processes in circumstellar disks is still 
incomplete. We know that molecular (collisional) viscosity is most certainly too low to be of 
practical interest. Turbulence driven by the magneto-rotational instability (MRI) 
is a most promising source of viscosity at present \citep{BH}, though other
mechanisms cannot be ruled out completely.

In this paper, we make no specific assumptions as to the source of viscosity and
define the coefficient of dynamical viscosity using the usual $\alpha$-prescription
of \citet{SS}
\begin{equation}
\mu=\alpha \, \Sigma \, \tilde{c}_s \, Z,
\end{equation} 
where spatially and temporally uniform $\alpha$ is set equal to 0.01 and 
$\tilde{c}_s=\sqrt{\partial {\cal P}/ \partial \Sigma}$ is the effective
sound speed. Our numerical simulations of embedded and T Tauri disks
indicate that $\alpha\simeq 0.001-0.01$ yields disk sizes and radial slopes 
that are in general agrement with observations \citep{VB4}. Lower values of $\alpha$ 
($ < 10^{-3}$) have little effect on the evolution of self-gravitating disks, 
whereas substantially higher values ($\ga 0.1$) quickly destroy the disks.
\citet{Hartmann98} also predicted similar values for $\alpha$ by
analyzing accretion rates in T Tauri disks.
Since viscosity in our  model is assumed to arise due to some physical processes 
{\it in the disk}, we keep $\alpha$ equal to zero during the early ``pre-disk'' phase of 
evolution and set $\alpha$ equal to 0.01 only when a circumstellar disk forms
around a central star.

\subsection{Initial condition}

The initial radial distributions of surface density $\Sigma$ and angular velocity $\Omega$ 
in our model cloud cores
are those characteristic of a collapsing axisymmetric magnetically
supercritical core \citep{Basu}
\begin{equation}
\Sigma={r_0 \Sigma_0 \over \sqrt{r^2+r_0^2}}\:,
\label{dens}
\end{equation}
\begin{equation}
\Omega=2\Omega_0 \left( {r_0\over r}\right)^2 \left[\sqrt{1+\left({r\over r_0}\right)^2
} -1\right],
\end{equation}
where $r_0$ is the radial scale length defined as $r_0 = k c_s^2 /(G\Sigma_0)$ and  $k= \sqrt{2}/\pi$.
These initial profiles are characterized by the important
dimensionless free parameter $\eta \equiv  \Omega_0^2r_0^2/c_s^2$
and have the property 
that the asymptotic ($r \gg r_0$) ratio of centrifugal to gravitational
acceleration has magnitude $\sqrt{2}\,\eta$ \citep[see][]{Basu}. 
The centrifugal radius of a mass shell initially located at radius $r$ is estimated to be
$r_{\rm cf} = j^2/(Gm) = \sqrt{2}\, \eta r$, where $j=\Omega r^2$ is the specific angular
 momentum. Since the enclosed mass $m$ is a linear function of $r$ at large radii,
this also means that $r_{\rm cf} \propto m$.
The gas has a mean molecular mass $2.33 \, m_{\rm H}$ and cloud cores are initially
isothermal with temperature $T=10$ K.

We present results from three sets of models, each with a different value of $\eta$.  
The standard model has $\eta = \eta_1 = 1.2 \times 10^{-3}$ based on typical values
$c_s = 0.19$ km s$^{-1}$, $\Sigma_0 = 0.12$~g~cm$^{-2}$, and
$\Omega_0 = 1.0$~km~s$^{-1}$~pc$^{-1}$. The outer radius is taken to
be $r_{\rm out} = 0.04$ pc, and the total cloud mass is $0.8\,\msun$. 
Other models with $\eta = \eta_1$ but different mass (outer radius) 
are generated by varying $r_0$ and $\Omega_0$ so that their product is constant. 
Note that, when $r_0$ is varied, $\Sigma_0$ has to be changed accordingly.
All clouds are characterized by the same ratio $r_{\rm out}/r_0\approx 6.0$. 
To generate the second set of models, $\eta = \eta_2 = 2.3 \times 10^{-3}$, we set
$\Omega_0 = 1.4$~km~s$^{-1}$~pc$^{-1}$ and all other quantities the
same as in the standard model with $\eta= \eta_1$. Models of
varying mass are then generated in the same manner as for the
$\eta_1$ models. The third set of models, with $\eta=\eta_3
= 3.4 \times 10^{-3}$, are also obtained in this way,
by first using $\Omega_0 = 1.7$~km~s$^{-1}$~pc$^{-1}$.
Overall, there are 7 models with $\eta = \eta_1$, 13 models with
$\eta = \eta_2 \simeq 2\, \eta_1$, 
and 12 with $\eta = \eta_3 \simeq 3\, \eta_1$.
The range of initial cloud masses ($M_{\rm cl}$) amongst our models is $0.3\,\msun-2.95\,\msun$.
The parameters of our models are listed in Table~\ref{table1} ($\eta_1$), Table~\ref{table2}
($\eta_2$), and Table~\ref{table3} ($\eta_3$).
We note that our model values of $\Omega_0 = (1.0-1.7)$~km~s$^{-1}$~pc$^{-1}$ are within a typical 
range of velocity gradients measured in dense starless cores by \citet{Caselli}.

\begin{table}
\begin{center}
\caption{Parameters of models with $\eta_1=1.2 \times 10^{-3}$
\label{table1}}
%\vspace{3 pt}
\begin{tabular}{clllll}
\hline\hline
Model & $r_0$ & $\Sigma_0$ & $\Omega_0$ & $r_{\rm out}$ & $M_{\rm cl}$  \\
\hline
 1 & 1209 & 0.13  & 1.14 & 7186  & 0.7 \\
 2 & 1382 & 0.12  & 1.0  & 8213  & 0.8 \\
 3 & 1728 & 0.093 & 0.8  & 10266 & 0.98 \\
 4 & 2074 & 0.077 & 0.67 & 12320 & 1.18 \\
 5 & 2937 & 0.055 & 0.47 & 17452 & 1.67 \\
 6 & 4147 & 0.039 & 0.33 & 24640 & 2.36 \\
 7 & 5184 & 0.031 & 0.27 & 30800 & 2.95 \\
 \hline
\end{tabular} 
\tablecomments{All distances are in AU, angular
velocities in km~s$^{-1}$~pc$^{-1}$, surface densities in g~cm$^{-2}$, and masses in $M_\odot$.}
\end{center}
\end{table} 

\begin{table}
\begin{center}
\caption{Parameters of models with $\eta_2=2.3 \times 10^{-3}$
\label{table2}}
%\vspace{3 pt}
\begin{tabular}{clllll}
\hline\hline
Model & $r_0$ &  $\Sigma_0$ & $\Omega_0$ & $r_{\rm out}$ & $M_{\rm cl}$  \\
\hline
 8 & 622  & 0.26  & 3.1  & 3696  & 0.35 \\
 9 & 691  & 0.23  & 2.8  & 4106  & 0.4 \\
 10 & 864  & 0.19  & 2.24 & 5133  & 0.5 \\
 11 & 1037 & 0.16  & 1.87 & 6160  & 0.6 \\
 12 & 1210 & 0.13  & 1.6  & 7187  & 0.7 \\
 13 & 1417 & 0.11  & 1.4  & 8213  & 0.8  \\
 14 & 1728 & 0.093 & 1.12 & 10267 & 0.98 \\
 15 & 2454 & 0.065 & 0.79 & 14579 & 1.4  \\
 16 & 2765 & 0.058 & 0.7  & 16428 & 1.57 \\
 17& 3456 & 0.046 & 0.56 & 20533 & 1.97 \\
 18& 3802 & 0.042 & 0.51 & 22587 & 2.16 \\
 19& 4147 & 0.038 & 0.47 & 24640 & 2.36 \\
 20& 4838 & 0.033 & 0.4  & 28747 & 2.75 \\
% 14& 5530 & 0.029 & 0.35 & 32853 & 3.15 \\
 \hline
\end{tabular} 
\tablecomments{All distances are in AU, angular
velocities in km~s$^{-1}$~pc$^{-1}$, surface densities in g~cm$^{-2}$, and masses in $M_\odot$.}
\end{center}
\end{table}

\begin{table}
\begin{center}
\caption{Parameters of models with $\eta_2=3.4 \times 10^{-3}$
\label{table3}}
%\vspace{3 pt}
\begin{tabular}{clllll}
\hline\hline
Model & $r_0$ & $\Sigma_0$ & $\Omega_0$ & $r_{\rm out}$ & $M_{\rm cl}$  \\
\hline
 21 & 518 & 0.31  & 4.53  & 3080  & 0.3 \\
 22 & 691  & 0.23  & 3.4  & 4106  & 0.4 \\
 23 & 864  & 0.19  & 2.7  & 5133  & 0.5 \\
 24 & 1037 & 0.16  & 2.26 & 6160  & 0.6 \\
 25 & 1210 & 0.13  & 1.94 & 7187  & 0.7 \\
 26 & 1417 & 0.11  & 1.7  & 8213  & 0.8  \\
 27 & 2073 & 0.077 & 1.13 & 12320 & 1.18 \\
 28 & 2420 & 0.066 & 0.97 & 14373 & 1.38 \\
 29 & 3283 & 0.049 & 0.72 & 19506 & 1.87  \\
 30 & 3802 & 0.042 & 0.62  & 22587 & 2.16 \\
 31 & 4147 & 0.039 & 0.57 & 24640 & 2.36 \\
 32 & 4493 & 0.036 & 0.52 & 26693 & 2.56 \\
% 32& 5530 & 0.029 & 0.43 & 32853 & 3.15 \\
% 13& 4838 & 0.033 & 0.4  & 28747 & 2.75 \\
% 14& 5530 & 0.029 & 0.35 & 32853 & 3.15 \\
 \hline
\end{tabular} 
\tablecomments{All distances are in AU, angular
velocities in km~s$^{-1}$~pc$^{-1}$, surface densities in g~cm$^{-2}$, and masses in $M_\odot$.}
\end{center}
\end{table}

\subsection{Numerical technique}
 
Hydrodynamic equations of the basic and viscous numerical models are solved in polar 
coordinates $(r, \phi)$ on a numerical grid with
$128 \times 128$ points. We use the method of finite differences with a time-explicit,
operator-split solution procedure. Advection is
performed using the second-order van Leer scheme.  The radial points are logarithmically spaced.
The innermost grid point is located at $r=5$~AU, and the size of the 
first adjacent cell lies in the range between 0.26~AU (model~21) 
and 0.36~AU (model~7). It means that the ratio $\Delta r/r$ of the cell size $\Delta r$ 
to radius $r$ is constant for a given cloud core and varies from $0.05$ (model~21) to 
$0.07$ (model~7). 
%It can be noticed from Fig.~\ref{figA1} that the aspect ratio $A=z/r$
%of the disk vertical scale height to radius is smaller than $\Delta r/r$ in the inner 70~AU, 
%which implies that a better radial (and angular) resolution is desirable in this inner region. 
%We have found that an increase in the resolution to $256 \times 256$ grid points makes 
%little influence on the accretion history and disk masses but helps save a considerable amount 
%of CPU time and, eventually, consider many more model cloud cores. We also point out that, due to
%the adopted log spacing in the radial direction, the resolution in the inner computational
%region is sufficient to fulfill the Truelove criterion in the disk.

 We introduce a ``sink cell'' at $r<5$~AU, 
which represents the central star plus some circumstellar disk material, 
and impose a free inflow inner boundary condition. About 95 per cent of 
the material crossing the inner boundary lands into the star, the
rest constitutes an inner circumstellar region of constant surface density.     
The dynamics of the inner region ($r<5$~AU) is not computed but it contributes 
to the total gravitational potential of the system. 
We do not account for a possible mass loss from the sink cell due to stellar jets, 
since our model is two-dimensional and
it is not clear how the mass ejection efficiency varies with stellar age. 
The outer boundary is such that the cloud has a constant mass and volume. 
More details on numerical techniques and relevant tests are given in \citet{VB2}.

\section{Masses of self-gravitating disks}
\label{SG}
In this section we present disk masses obtained using the basic numerical approach
described in \S~\ref{basic}. In the framework of this numerical model, disk masses 
are controlled by the rate of mass infall from 
the surrounding envelope and the radial transport of mass and angular momentum 
due to gravitational torques. No viscous torques are present in this case.

An accurate determination of disk masses in numerical simulations of collapsing cloud cores is not
a trivial task. Self-consistently formed circumstellar disks have a wide range of masses 
and sizes, which are not known {\it a priori}. However, numerical and observational 
studies of circumstellar disks indicate that the disk
surface density is a declining function of radius. Therefore, we distinguish
between disks and infalling envelopes using a critical surface density for the 
disk-to-envelope transition, for which we choose a value of $\Sigma_{\rm tr}=0.1$~g~cm$^{-2}$.
This choice is dictated by the fact that densest {\it starless} cores have surface
densities only slightly lower than the adopted value of $\Sigma_{\rm tr}$.
In addition, our numerical simulations indicate that self-gravitating disks have 
sharp outer edges and the gas densities of order $0.01-0.1$~g~cm$^{-2}$ 
characterize a typical transition region between the disk and envelope \citep{VB3}.

To compare disk masses in three distinct phases of stellar evolution
we need an evolutionary indicator to distinguish between Class 0, Class I, 
and Class II phases. 
We use a classification of \citet{Andre}, who
suggest that the transition between Class 0 and Class I objects occurs when
about $50\%$ of the initial cloud core is accreted onto the protostar/disk system.
The Class II phase is consequently defined by the time when the
infalling envelope clears and its total mass drops below 10\% of the initial cloud 
core mass $M_{\rm cl}$.
It should be mentioned here that there exists no unique classification scheme for protostars.
For instance, \citet{VB0} have proposed a classification scheme that hinges 
on a temporal behaviour 
of bolometric luminosity $L_{\rm bol}$. They identify Class~0 phase with a period of 
temporally increasing $L_{\rm bol}$ and Class~I phase with a later period of decreasing 
$L_{\rm bol}$. The peak in $L_{\rm bol}$ corresponds to the evolutionary time when 
$50 \pm 15$ per cent of the cloud core mass has been accreted by the protostar.
Observers prefer to classify protostars by their spectral energy distributions. 
For instance, \citet{Lada87} and \citet{Andrews} use the values of power-law index $n$ 
(defined by $\nu F_{\nu} \propto \nu^n$, where $F_{\nu}$ is the infrared flux density at frequency 
$\nu$) to distinguish between Class~I and Class~II objects. 
\citet{Andre} use the ratio of submillimeter to bolometric luminosity
$L_{\rm submm}/L_{\rm bol}$ for the same purposes. Although these classifications are
physically related, it is possible that they may differ from our adopted classification, 
especially if some of the gas is removed
from the cloud core and does not accrete. We acknowledge that the difference 
in the existing classification schemes may systematically shift our results.

\subsection{Temporal evolution of self-gravitating disks}
\label{tempevol}
\begin{figure}
  \resizebox{\hsize}{!}{\includegraphics{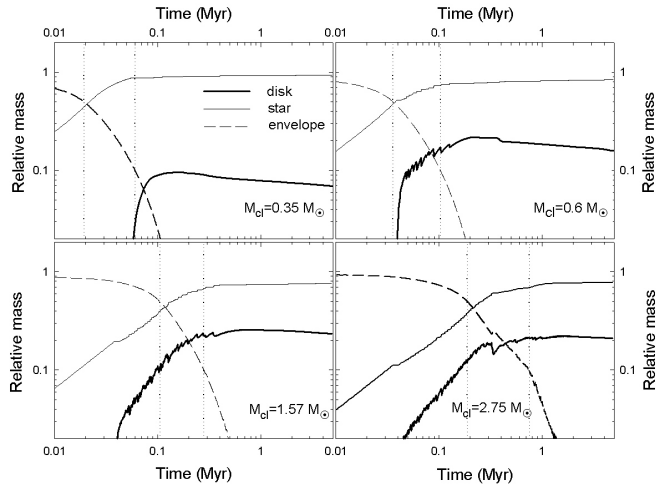}}
% \centering
%  \includegraphics{figure1.eps}
      \caption{Disk masses (thick solid lines), stellar masses (thin solid lines),
      and envelope masses (dashed lines) obtained in the framework of 
      the basic numerical approach (see \S~\ref{basic}) 
      for model~8 (upper left), model~11 (upper right), 
      model~16 (lower left), and model~20 (lower right).
      All masses are calculated relative to the corresponding 
      initial cloud core mass $M_{\rm cl}$. The horizontal axis shows time elapsed since the 
      formation of a central star. The left/right vertical dotted lines mark the onset 
      of Class I/II phases, respectively.}
         \label{fig1}
\end{figure}

We start by comparing the long-term evolution of disk masses in four sample models, which are 
chosen to represent a full spectrum of the initial cloud core 
masses with a constant ratio $\eta=2.3\times 10^{-3}$ of centrifugal to gravitational 
acceleration. Figure~\ref{fig1} shows the disk masses (thick solid lines), stellar masses (thin
solid lines), and envelope masses (dashed lines) in model~8 (upper left, $M_{\rm cl}=0.35~M_\odot$),
model~11 (upper right, $M_{\rm cl}=0.6~M_\odot$), model~16 (lower left, $M_{\rm cl}=1.57~M_\odot$),
and model~20 (lower right, $M_{\rm cl}=2.75~M_\odot$). The horizontal axis is the time elapsed 
since the formation of a central star.
All masses are calculated relative to the corresponding 
initial cloud core mass $M_{\rm cl}$. The vertical dotted lines mark the onset 
of Class I (left)  and Class~II (right) phases.

Our numerical simulations demonstrate that cloud cores with constant $\eta$ (but
different size) form disks roughly at the same physical time after the formation 
of a central star but in distinct stellar evolution phases. For instance,
model~8 starts to build a disk in the late Class I phase, while model~20 does that in the midst
of Class 0 phase. Cloud cores of greater mass tend to form disks in the earlier phase of stellar 
evolution than their low-mass counterparts. 

The upper panel of Fig.~\ref{fig1} identifies cases when Class~0 stars have no disks
associated with them.  
We emphasize here that due to the use of the sink cell in our numerical code
we can resolve only those disks whose outer radii are larger than 5~AU. It means that even though 
circumstellar disks do not form around some Class 0 objects in our numerical simulations, 
as in models~8 and 11 
(top panels in Fig.~\ref{fig3}), one may suppose that such disks still exist  
but their size is simply smaller than that of the sink cell. 
Our test runs with a sink cell set to 0.5~AU (instead of 5~AU) confirm
that disks indeed forms earlier in the evolution but quickly expands to 5~AU and beyond.
Hence, the absence of disks around some Class~0 objects may be to some extent 
caused by a finite-size sink cell. However, it is still possible that some Class~0 
objects have no disks associated with them. This is particularly true for those objects that 
develop from cloud cores with low rates of rotation, since in this case
a (substantial) portion of the disk material will be characterized
by the centrifugal radius (estimated as $r_{\rm cf}=j^2/(Gm)$) that is
smaller than the radius of the protostar, $4~R_\odot$.
A numerical code with realistic treatment of the protostar formation is needed
to accurately address this issue.

Another interesting feature of self-gravitating disks that can be seen in Fig.~\ref{fig1} 
is that the disk-to-star mass 
ratio never exceeds some characteristic value, approximately $0.35-0.4$, irrespective of the
initial cloud core mass. This is rather counterintuitive. For instance, models~16 and 20 (bottom panels
in Fig.~\ref{fig1}) form disks in the Class 0 phase, which is characterized by envelope 
masses that are considerably greater than those of the protostars. As a consequence,
one may expect the formation of a disk with mass that is at least comparable 
to or even greater than that of a protostar. It turns out, however, that circumstellar disks 
that form in the {\it Class~0 or early Class~I phase} around stars with mass 
$M_\ast \ga 0.6~M_\odot$ develop vigourous gravitational 
instability -- a very efficient means of inward mass transport 
that helps keep the disk mass well below that of the protostar. 
Sharp drops in the disk mass (or equivalent surges in the stellar mass) 
seen in the upper right and bottom panels of Figure~\ref{fig1} are a manifestation of 
this process \citep[see e.g.][]{VB1,VB2}.  
\citet{Kratter} also predict that disks around stars with mass greater than $1.0~M_\odot$
are expected to be vigorously gravitationally unstable in the early embedded phase of
disk evolution. 
The late evolution phase (Class II) sees only an insignificant decline of disk mass with time.

\begin{figure}
  \resizebox{\hsize}{!}{\includegraphics{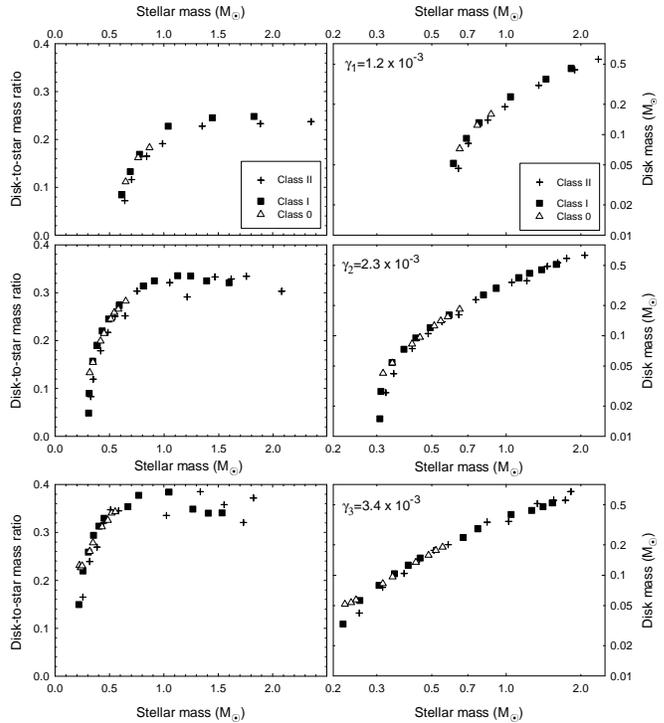}}
      \caption{Time-averaged disk masses (right column) and
      time-averaged disk-to-star mass ratios (left column) versus time-averaged stellar masses. 
      Open triangles, filled squares, and plus signs show the data for Class~0, Class~I, 
      and Class~II 
      systems, respectively. In particular, top panels show the data for models 
      with $\eta_1 = 1.2 \times 10^{-3}$, whereas middle and bottom panels show the data for 
      models with $\eta_2\simeq 2\eta_1$ and $\eta_3\simeq 3\eta_1$, respectively.}
         \label{fig2}
\end{figure}

\subsection{Disk masses versus stellar masses}
We use 32 models, the parameters of which are listed in 
Tables~\ref{table1}-\ref{table3}, to analyse the statistical relations between time-averaged 
disk and stellar masses in the Class 0/I/II evolution phases. 
Time averaging is performed separately for each evolution phase over the duration of the phase.
The age of the oldest disk in our sample 
is about 3~Myr. Since disk masses in the Class II phase have a tendency to decline with 
time and the actual
lifetime of Class II objects may be longer, we expect that
we might have somewhat overestimated disk masses around Class II objects. This, obviously, does not
effect our estimates of Class 0/I disk masses. 

Figure~\ref{fig2} shows time-averaged disk masses (right column) and time-averaged 
disk-to-star mass ratios (left column) versus time-averaged stellar masses 
for Class~0 (open triangles), Class~I (filled squares) and Class~II  objects (plus signs).
 Several interesting conclusions can be drawn 
by analysing the figure. 
\begin{enumerate}
 \item Time-averaged disk and stellar masses ($\langle M_{\rm d}\rangle$ and 
 $\langle M_\ast \rangle$, 
 respectively) in models with the same ratio of rotational to gravitational 
acceleration fall onto a unique evolutionary track in the $\langle M_{\rm d} \rangle - \langle M_\ast
\rangle$  diagram. 
\item  Class 0 objects occupy the lower-left part of each evolutionary track. No stars with 
$\langle M_\ast \rangle \ga 0.9~M_\odot$ have Class 0 disks (but they have Class~I/II disks). 
In other words, only low-mass stars (within our range of interest, 
$0.2~M_\odot \la \langle M_\ast \rangle  \la 2.0~M_\odot$) can harbour Class~0 disks.
\item Stars of {\it equal mass} have disks with similar masses, regardless of the stellar 
evolution phases. 
\item In each stellar evolution phase, disk-to-star mass ratios tend to have greater values 
for stars of greater mass. However, there is a clear saturation effect for stars with
masses greater than $1.0~M_\odot$ -- disk masses never grow above $40\%$ 
of the stellar masses, even for models with the largest values of $\eta = \eta_3=3.4\times 10^{-3}$.
The saturation of disk-to-star mass ratios is caused by the onset of vigorous
gravitational instability in circumstellar disks that form in the Class~0 and early Class~I phase.
\end{enumerate}

\begin{table*}
\begin{center}
\caption{Summary of disk properties}
\label{table4}
%\vspace{3 pt}
\begin{tabular}{ccccccc}
\hline
\hline
disk type & $ \!\!\! \overline{M}_{\rm d,C0}$ & $\!\!\! \overline{M}_{\rm d,CI}$ & $\!\!\! \overline{M}_{\rm d,CII}$ & 
$\!\!\! \left( \langle M^{\rm min}_{\rm d,C0}\rangle : \langle M^{\rm max}_{\rm d,C0} \rangle \right)$ & 
$ \!\!\! \left( \langle M^{\rm min}_{\rm d,CI} \rangle : \langle M^{\rm max}_{\rm d,CI} \rangle \right) $
 &  $ \!\!\! \left( \langle M^{\rm min}_{\rm d,CII} \rangle : \langle M^{\rm max}_{\rm d,CII} 
 \rangle \right)$ \\
\hline
self-gravitating& 0.09 & 0.10 & 0.12  & $(0.04 : 0.2)$ & $(0.02 : 0.5)$  & $(0.03 : 0.7)$ \\
viscous & 0.10 & 0.11 & 0.06  & $(0.05 : 0.19)$  & $(0.02 : 0.55)$  & $(0.004 : 0.6)$ \\
 \hline
\end{tabular} 
\tablecomments{Mean masses $\overline{M}_{\rm d,C0}$, $\overline{M}_{\rm d,CI}$, and 
$\overline{M}_{\rm d,CII}$ of Class 0, Class I, and Class II disks, respectively, are
calculated according to equation~(\ref{mean}). Angle brackets $\langle \, \rangle$ denote
time averaging over the duration of a particular stellar evolution phase. 
Indices ``min'' and ``max'' refer to minimum and maximum time-averaged disk masses, respectively. 
All disk masses are in $M_\odot$.}
\end{center}
\end{table*}

We summarize the main properties of our model self-gravitating disks in Table~\ref{table4}.
In order to facilitate the comparison of our numerical results with 
observations, we calculate mean disk masses (in each evolution phase) 
by weighting the time-averaged disk masses according to the corresponding stellar masses.
The relative importance of the stellar masses is calculated using the following power-law 
weight function
\begin{equation}
{\cal F}_{\rm K}\left(\langle M_\ast \rangle\right) = \left\{ \begin{array}{ll} 
   A \, \langle M_\ast \rangle^{-1.3} & \,\, 
   \mbox{if $\langle M_\ast \rangle \le 0.5~M_\odot$ } \\ 
   B \, \langle M_\ast \rangle^{-2.3} & \,\, 
   \mbox{if $\langle M_\ast \rangle > 0.5~M_\odot$ }.  \end{array} 
   \right. 
   \label{function}  
 \end{equation}
The slope of ${\cal F}_{\rm K}\left(\langle M_\ast \rangle\right)$ is typical for 
the Kroupa initial mass function (IMF) of stars \citep{Kroupa01}. 
The mean disk masses are then calculated as follows
\begin{equation}
\overline{M}_{\rm \rm d,ph} = { \sum^N_{i=1} \langle M^i_{\rm d,ph} \rangle \, \,
{\cal F}_{\rm K}\left(\langle M^i_{\rm \ast,ph} \rangle\right) \over 
\sum^N_{i=1} {\cal F}_{\rm K} \left(\langle M^i_{\rm \ast,ph} \rangle\right) },
\label{mean}
\end{equation}
where $\langle M^i_{\rm d,ph} \rangle$ and $\langle M^i_{\rm \ast,ph} \rangle$ are the 
time-averaged disk and stellar masses in the $i$-th model (see Fig.~\ref{fig2})  
and index ``ph'' stands for Class 0 (C0), Class I (CI), or 
Class II (CII) stellar evolution phases. The summation is performed separately for each 
evolution phase. 
In particular, there are $N=20$ models that have Class~0 disks and $N=31$ models that have
Class I disks. All 32 models have Class II disks, of course. 
The ratio of the  normalization constants $A/B=2$ needed to evaluate 
equation~(\ref{mean}) is found using the continuity condition 
at $\langle M_\ast\rangle=0.5~M_\odot$.

%The slope of the initial mass function
%of stars is set to 1.3 in the $0.08-0.5~M_\odot$ range and 2.3 for stars more massive than 
%$0.5~M_\odot$ \citep{Kroupa01}. 

Table~\ref{table4} indicates that the mean masses slightly increase along the stellar evolution
sequence from $\overline{M}_{\rm d,C0}=0.09~M_\odot$ around Class~0 objects to 
$\overline{M}_{\rm d,CII}=0.12~M_\odot$ around Class~II objects.
The minimum and maximum time-averaged disk masses 
($\langle M^{\rm min}_{\rm d,ph} \rangle$ and $\langle M^{\rm max}_{\rm d,ph} \rangle$, 
respectively) 
in each evolution phase are shown in the last three columns of Table~\ref{table4}. It is 
seen that both the Class I and Class II disks have a similar range of masses, while 
Class 0 disks have a somewhat narrower range. The smallest time-averaged mass of a 
self-gravitating disk found in our numerical simulations is $\langle M_{\rm d} \rangle = 0.02~M_\odot$.

\section{Masses of viscous disks}
\label{VD}
In this section we present disk masses obtained in the framework of the viscous numerical approach
described in \S~\ref{viscmodel}. We seek to determine the effect of viscosity,
and associated viscous torques, on the masses of self-consistently formed circumstellar disks.
Our comparison study of viscous versus self-gravitating disks have shown that
viscous disks may lack a sharp transition boundary between the disk and the envelope \citep{VB4}.
This smearing of the disk's outer physical boundary is particularly pronounced in 
disks of low mass and in the late Class~II phase.
In this situation, it is somewhat arbitrary to adopt a value of $\Sigma_{\rm tr}=0.1$~g~cm$^{-2}$
for the disk-to-envelope transition. Nevertheless, we have recalculated disk
masses with $\Sigma_{\rm tr}=0.01$~g~cm$^{-2}$ and found that this can increase 
masses of Class~II disks by $25\%$ at most. Class 0/I disks are unaffected by the value of 
$\Sigma_{\rm tr}$, since they usually have sharp outer physical boundaries.

\begin{figure}
  \resizebox{\hsize}{!}{\includegraphics{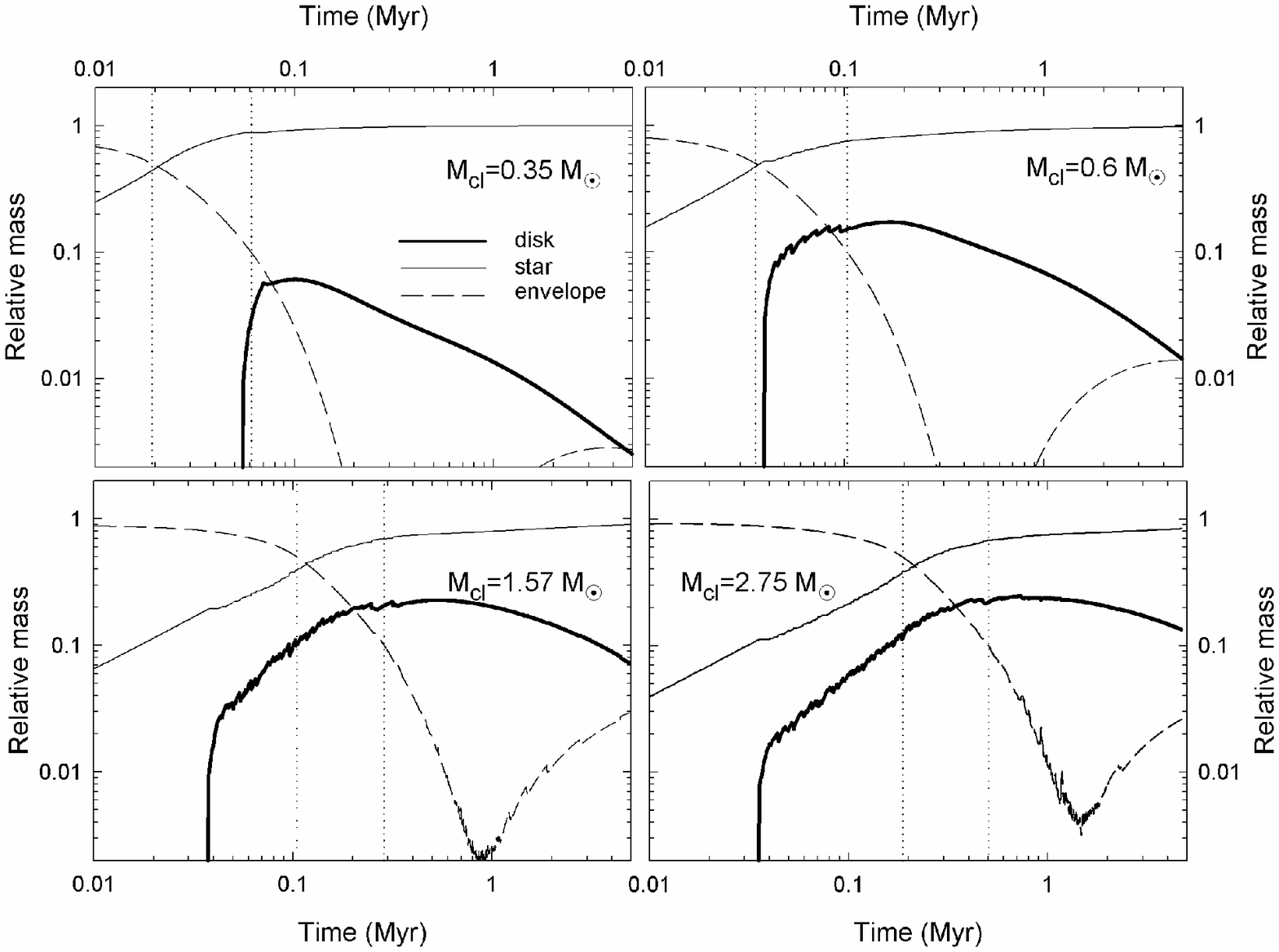}}
      \caption{The same as Figure~\ref{fig1} but obtained in the framework of the viscous 
      numerical approach described in \S~\ref{viscmodel}}
         \label{fig3}
\end{figure}

\subsection{Temporal evolution of viscous disks}
In order to investigate the temporal evolution of viscous disks, we consider the same 
four sample model cloud cores as in \S~\ref{tempevol}. Figure~\ref{fig3}
presents  disk masses (thick solid lines), stellar masses (thin solid lines),
and envelope masses (dashed lines) in model~8 (upper left), model~11 (upper right), model~16 (lower
left), and model~20 (lower right). The horizontal axis is time elapsed since the formation of a central
star. All masses are calculated relative to the corresponding 
initial cloud core mass $M_{\rm cl}$.
The vertical dotted lines mark the onset of Class I phase (left line) and Class II phase (right line).

The comparison of Figures~\ref{fig1} and \ref{fig3} reveals a major difference between self-gravitating
and viscous disks  -- the latter have 
considerably smaller masses in the {\it late} Class II phase than the former. 
For instance, model~8 in Figure~\ref{fig1} has a final ($t=5$~Myr) stellar mass 
$M_\ast = 0.33~M_\odot$ and disk mass   
$M_{\rm d}=2.5\times10^{-2}~M_\odot$, while the same model in Figure~\ref{fig3} has  
$M_{\rm d}=8.5\times 10^{-4}~M_\odot$.
Such a large contrast, however, diminishes for stars of greater mass. For instance, model~16
in Figure~\ref{fig1} has $M_\ast=1.57~M_\odot$ and $M_{\rm d}=0.37~M_\odot$, while the same model 
in Figure~\ref{fig3} has $M_{\rm d}=0.11~M_\odot$.
On the other hand, both the viscous and self-gravitating disks have comparable masses in the 
Class 0/I phases. Even the early Class II phase sees little difference between the masses 
of viscous and self-gravitating disks, provided that the disks are formed from 
molecular cloud cores of similar mass.
Figure~\ref{fig3} indicates that a small portion of the viscous disk material 
returns to the envelope in the late disk evolution. This is because we have imposed a fixed 
gas column density threshold for the disk-to-envelope transition, but
viscous disks expand in the late evolution due to the action of viscous torques, 
which are in fact positive in the outer disk regions \citep{VB4}. This numerical 
effect may slightly reduce the resulted viscous disk masses in the Class II phase.

\begin{figure}
  \resizebox{\hsize}{!}{\includegraphics{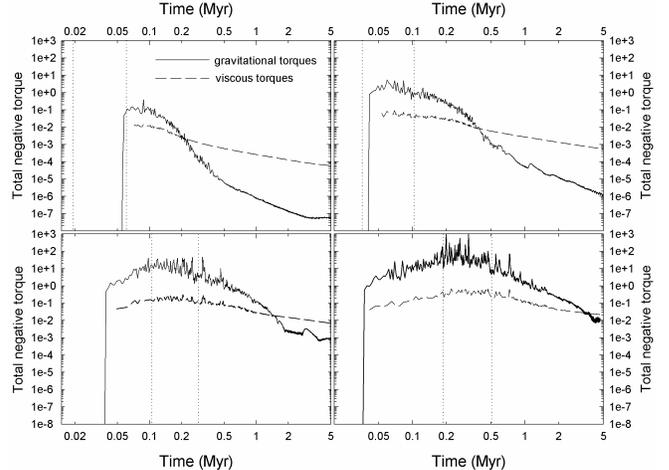}}
      \caption{Total negative gravitational torques (solid lines) and total negative
      viscous torques (dashed lines) versus time elapsed since the formation of a central star
      in model~8 (upper left), model~11 (upper right), model~16 (lower left), and model~20 
      (lower right). The vertical dotted lines mark the onset of Class I~phase (left line) 
      and Class~II phase (right line). The torques are in units of $8.7\times 10^{40}$
      ~g~cm$^{2}$~s$^{-2}$.}
         \label{fig4}
\end{figure}

Why does the late stellar evolution phase reveal such a noticeable difference in the 
disk masses (and, consequently, in disk-to-star mass ratios) between viscous and self-gravitating 
disks and why is there little difference in the earlier phases? The reason for that can be understood
if we consider the temporal evolution of viscous and gravitational torques in the disk.
Figure~\ref{fig4} shows total negative torques due to disk self-gravity (${\cal T}_{\rm g}$,
solid lines) and viscosity (${\cal T}_{\rm v}$, dashed lines) in our four typical models:
model~8 (upper left), model~11 (upper right), model~16 (lower left), and model~20 (lower right).
The horizontal axis shows time elapsed since the formation of a central star.
The total negative torque is computed by summing up local negative torques $\tau(r,\phi)$ 
in each computational zone occupied by the disk. In particular, ${\cal T}_{\rm g}$ is computed as 
a sum of $\tau_{\rm g}(r,\phi) = -m(r,\phi) \, \partial \Phi(r,\phi) / \partial \phi$, where $m(r,\phi)$
and  $\Phi(r,\phi)$ are the gas mass and gravitational potential in a cell with polar coordinates $(r,\phi)$,
respectively. The total negative viscous torque is calculated in a similar manner by summing up 
all local negative viscous torques $\tau_{\rm v}(r,\phi)=r\, (\nabla \cdot \mathbf{\Pi})_\phi  
\, S(r,\phi)$, where $S(r,\phi)$ is the surface area occupied by a cell with polar coordinates ($r,\phi$).
The local torques $\tau_{g}(r,\phi)$ and $\tau_{\rm v}(r,\phi)$ are actually the local 
azimuthal components of the corresponding 
forces, acting on a fluid element with mass $m(r,\phi)$, multiplied by the arm $r$.

Figure~\ref{fig4} indicates that gravitational torques always prevail over viscous 
torques in the Class 0 and
Class I phases. As a result, the disk mass in these phases is determined by
the interplay between the mass load from an infalling envelope and the rate of mass
and angular momentum transport in the disk due to gravitational torques. In the Class II phase,
the strength of gravitational torques diminishes, partly due to a stabilizing influence of a growing
central star and partly due to the exhausted mass reservoir in the infalling envelope.
As a result, viscous torques, the strength of which shows only a mild decline with time, 
take a leading role in the Class II phase. They act to further decrease the disk mass when 
gravitational torques fail to do so. 

The actual time in the Class II phase when ${\cal T}_{\rm v}$ become greater than
${\cal T}_{\rm g}$ is distinct for models with different initial cloud core masses (but similar 
ratios of rotational to gravitational acceleration $\eta$). 
Low-mass cloud cores form disks late in the evolution (see e.g. model~8 in Fig.~\ref{fig3}).
The resulting disks have low masses, merely due to the fact that most of the cloud core material 
has already been accreted directly onto the protostar in the pre-disk phase. As a result, 
gravitational instability is underdeveloped and viscous torques quickly take over the 
gravitational ones in the beginning of the Class II phase. 
Conversely, high-mass cloud cores form disks early in the evolution 
(see e.g. model~16 and 20 in Fig.~\ref{fig3}). The resulting disks are characterized 
by considerable masses. As a consequence, gravitational torques prevail over viscous torques 
throughout a considerable portion of the Class II phase. The interested reader 
is referred to \citet{VB5,VB4} for a detailed discussion on gravitational torques and
appearances of self-gravitating and viscous disks.

\subsection{Relation between disk and stellar masses}
\begin{figure}
  \resizebox{\hsize}{!}{\includegraphics{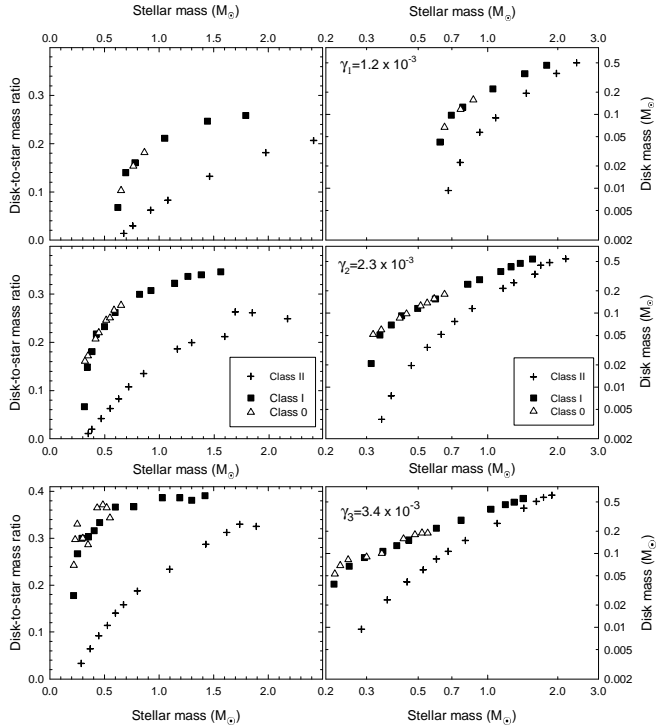}}
      \caption{The same as Figure~\ref{fig2} only obtained in the framework of the viscous numerical
      approach described in \S~\ref{viscmodel}.}
         \label{fig5}
\end{figure}

In this section we derive statistical relations between time-averaged disk masses 
($\langle M_{\rm d}\rangle$), time-averaged stellar masses ($\langle M_\ast \rangle$) , 
and time-averaged disk-to-star mass ratios ($\langle \xi \rangle$) for viscous disks 
in each stellar evolution phase. Time averaging is performed separately in each evolution phase 
over the duration of the phase.
Figure~\ref{fig5} shows $\langle M_{\rm d}\rangle$ (right column) and $\langle \xi \rangle$  
(left column) versus $\langle M_\ast \rangle$ for Class 0 objects (open triangles),
Class I objects (filled squares) and Class II objects (plus signs).
The top-to-bottom rows show the data for models with $\eta_1=1.2\times 10^{-3}$, 
$\eta_2=2.3\times 10^{-3}$, and $\eta_3=3.4\times 10^{-3}$, respectively.

The comparison of Figures~\ref{fig2} and \ref{fig5} reveals a distinct feature of viscous disks
around objects of equal stellar mass -- Class II disk have systematically lower masses than their 
Class 0/I counterparts. 
This is in sharp contrast to self-gravitating disks studied in \S~\ref{SG} where
we have found that Class O/I/II objects with equal stellar mass 
possess self-gravitating disks of essentially similar mass (see Fig.~\ref{fig2}). 
The difference in masses between Class II and Class 0/I viscous disks 
is considerable for low-mass stars (up to a factor of 10), but it diminishes for stars 
of greater mass. However, we have to keep in mind that low-mass stars are statistically more
important than their high-mass counterparts.

The dependence of $\langle \xi \rangle$ on $\langle M_\ast \rangle$
in Figure~\ref{fig5} is similar to that of Figure~\ref{fig2} -- stars of greater mass tend 
to have greater disk-to-star mass ratios. However, as for the case of disk masses, 
Class II viscous disks have disk-to-star mass ratios that are lower than those 
of self-gravitating disks. For instance, the smallest $\langle \xi \rangle$ in self-gravitation disks
(Fig.~\ref{fig2}) is 0.05, while viscous disks have $\langle \xi \rangle$ as small as 0.01.

{It is interesting to compare our results with the recent work of \citet{Kratter}, who use 
semi-analytical models to describe the time evolution of embedded, accreting protostellar disks.
As in our model, they account for disk self-gravity and MRI-induced viscosity but
also include the effect of stellar and envelope irradiation. They obtain maximum disk-to-star
mass ratios in the Class~I phases of order $30\% - 40\%$, in excellent agreement with our results.
The disk mass in their fiducial $1~M_\odot$ model shows a temporal behaviour similar to that 
shown in the lower-left panel of Fig.~\ref{fig3} -- it grows in the Class~0 phase,
reaches a maximum in Class~I phase, and declines afterwards.  
They also predict that higher mass stars have relatively more massive disks and this trend
is expected to extend to stellar masses much greater than those considered in our work ($\la 2.0~M_\odot$).

We calculate the mean masses of viscous disks using equation~(\ref{mean})
and summarize the main properties of our viscous disks in Table~\ref{table4}.
It is evident that the range of time-averaged masses and the values of mean masses are quite similar
in both the self-gravitating and viscous disks. The latter tend to have slightly larger
mean disk masses in the Class 0/I phases, which is explained by the fact that viscous
torques tend to oppose gravitational ones in the early disk evolution \citep{VB4}.
The only significant difference comes from
Class~II viscous disks, which have a factor of two lower mean mass, 
$\overline{M}_{\rm CII}=0.06~M_\odot$, than that of self-gravitating disks.
Class II viscous disks also feature the lowest time-averaged disk mass found in our numerical
simulations, $\langle M^{\rm min}_{\rm d,CII} \rangle=0.004~M_\odot$.

Finally, we seek to determine the relation between time-averaged disk and stellar masses 
in each stellar evolution phase separately and for the three phases taken altogether.
%for all model cloud cores of our sample, irrespective of their initial values of $\eta$.
Figure~\ref{fig6} shows $\langle M_{\rm d} \rangle$ versus $\langle M_\ast \rangle$
for our 32 model cloud cores. 
In particular, the open triangles, filled squares, and plus signs 
show the data for the Class 0, Class I, and Class II phases, respectively. 
It is evident that the upper-mass stars have a considerably narrower scatter in disk masses than 
the lower-mass stars. The least-squares best fit to {\it all} data in Figure~\ref{fig6} 
(solid line) yields the following relation between time-averaged stellar and disk masses 
(in $M_\odot$) for the three phases taken altogether
\begin{equation}
\langle M_{\rm d} \rangle = \left( 0.2\pm 0.05 \right) \, \langle M_\ast \rangle^{1.3\pm 0.15}.
\label{rel1}
\end{equation}

The relation between time-averaged disk and stellar masses in each stellar evolution phase
is, however, quite different from that expressed by equation~(\ref{rel1}).
The least-squares best fits performed separately for each evolution phase yield the 
following relations
\begin{eqnarray}
\label{rel3}
\langle M_{\rm d,CO} \rangle &=& \left( 0.2\pm 0.05 \right) \, \langle M_{\rm \ast,C0} \rangle^{0.7\pm 0.2}, \\
\langle M_{\rm d,CI} \rangle &=& \left( 0.3\pm 0.04 \right) \, \langle M_{\rm \ast,CI} \rangle^{1.3\pm0.15}, \\
\langle M_{\rm d,CII} \rangle &=& \left( 0.2\pm 0.08 \right) \, \langle M_{\rm \ast,CII} \rangle^{2.2\pm0.2}, 
\label{rel5}
\end{eqnarray}
where indices CO, CI, and CII refer to the Class~0, Class~I, and Class~II phases, respectively.
The dependence of disk masses on stellar masses becomes progressively steeper
along the sequence of stellar evolution phases. 

The above relations between time-averaged stellar and disk masses were derived for 
viscous disks, in which both viscosity and self-gravity are 
at work. Purely self-gravitating disks (zero viscosity) have a similar relation 
between time-averaged disk and stellar masses when {\it all} stellar evolution phases 
are considered altogether, namely equation~(\ref{rel1}).
However, when each phase is taken separately, the exponents are different from those derived
for viscous disks in equations~(\ref{rel3})-(\ref{rel5}). To avoid confusion, we provide here 
the relations for viscous disks only, since we believe that some sort of viscosity should 
be present in circumstellar disks.

\begin{figure}
  \resizebox{\hsize}{!}{\includegraphics{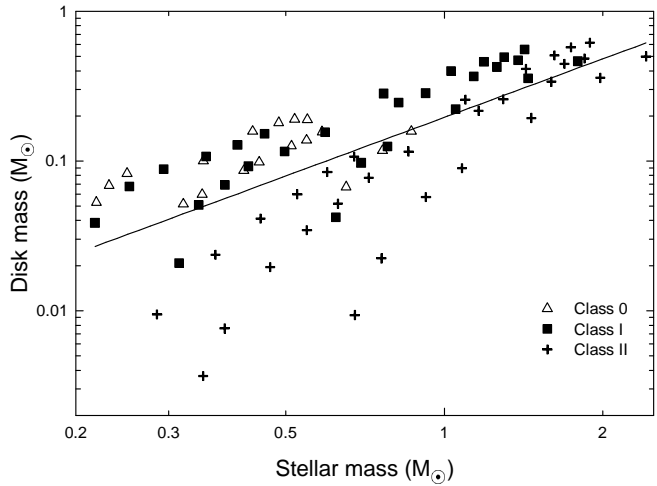}}
      \caption{Time-averaged disk masses versus time-averaged stellar masses for 32~models of our sample
      (see Tables~\ref{table1}-\ref{table3}). Open triangles, filled squares, and plus signs
      show the data for Class~0, Class~I, and Class~II systems, respectively. The solid line gives
      a least-squares best fit to all data in the plot. }
         \label{fig6}
\end{figure}

Observations provide conflicting evidence as to the relation between disk and stellar masses
in YSOs. For instance, \citet{Natta01} found a marginal correlation between disk 
and stellar masses, albeit with a substantial dispersion. On the other hand, \citet{Andrews}
claimed no correlation. The correlation in Figure~\ref{fig6} can be broken if 
a large population of objects occupies both the upper-left and lower-right portions of the 
$\langle M_{\rm d} \rangle - \langle M_\ast \rangle$ diagram. While the latter is feasible due to
the incompleteness of our sample of model cloud cores (we have not considered sufficiently low 
values of the ratio $\eta$), the former is unlikely due to the saturation of disk masses 
discussed in the context of Figure~\ref{fig2} and earlier considered by \citet{Adams89} and
\citet{Shu90} . No stars can harbour disks with masses comparable to or greater than 
the stellar mass! 
It is likely that the uncertainty in disk measurements may introduce a considerable scatter
to the observational data which precludes the correlation analysis.

\section{Discussion}
\label{discuss}

There exist a growing concern that disk masses around YSOs
may be systematically underestimated by conventional observational methods.
The simplest argument in favour of this conjecture is a high frequency of detection
of Jupiter-like exosolar planets. Both the core accretion and gravitational instability 
models require disk masses of order $0.1~M_\odot$ for Jupiter-like planets to form, 
which is about an order of magnitude 
higher than median disk masses inferred by \citet{Andrews} (hereafter, AW).
Another argument in favour of fairly massive disks is the fact that resolved disks 
show rich non-axisymmetric structures, such as {\it multiple} spiral arms or arcs in 
the case of AB Aurigae \citep{Fukagawa, Grady99} and HD 100546 \citep{Grady01}. Such structures 
are difficult to sustain in low-mass disks. To remind the reader, a companion star or
giant planet would  most likely trigger a two-armed spiral response in the disk of a target and 
not a flocculent multi-arm structure. 

As was discussed in the Introduction, measurements of disk masses suffer from the uncertainties
in dust opacities and disk radial structure, which could lead to a substantial underestimate
of disk masses \citep[e.g.][]{Hartmann}. Alternatively, the disk may mask itself, especially late in
the evolution, by locking a significant
amount of its mass in the form of large dust grains or planetesimals which are inefficient emitters
at submillimeter wavelengths \citep{Andrews}. 

Our numerical modeling supports the supposition of \citet{Hartmann} that circumstellar disks are 
more massive than currently inferred from observations. The mean masses of disks 
in our models range between $0.10-0.11~M_\odot$ for Class~I objects and $0.06-0.12~M_\odot$
for Class II objects, well in excess of the AW's median masses for either 
Class I disks ($3\times 10^{-2}~M_\odot$) or, especially, Class II disks ($3\times 10^{-3}~M_\odot$).
Much the same, our Class 0 disks have mean masses $0.09-0.10~M_\odot$, while 
five Class 0 disks in a sample of \citet{Brown} have a mean mass of order $0.01~M_\odot$,
though the statistics in the case of Class~0 objects is inadequate to draw any firm conclusions.

A closer look at the data provided by AW reveals that our numerical modeling fails to
reproduce only the lower bound on the measured disk masses but succeeds at reproducing the
upper bound. For instance, figure~15 of AW
indicates that Class I disk masses lie in the range $M_{\rm d,CI}=0.001-0.6~M_\odot$,
while our Class I disks have masses in the range 
$\langle M_{\rm d,CI} \rangle =0.02-0.55~M_\odot$. 
Similarly, AW's Class II disk masses lie in the range $M_{\rm d,CII}=5\times 
10^{-4} - 0.5~M_\odot$, while our disks have masses in the range 
$\langle M_{\rm d,CII} \rangle = 4\times 10^{-3}-0.7~M_\odot$.
If disk masses are indeed systematically underestimated by conventional observational methods, then
low-mass disks should suffer from this problem to a greater extent than do upper-mass disks.
The reason for this is not well understood.

It is worth noting that our Class 0 disk have masses that are similar to those 
of Class I disks. This is in agreement with the modeling of circumstellar
envelope dust emission of 6 deeply embedded systems by \citet{Looney}, who have found 
that Class 0 systems do not 
have disks that are significantly more massive than those in Class I systems.
This led them to conclude that the disk in Class 0 systems must quickly and
efficiently process $\sim 1.0~M_\odot$ of material from the envelope onto the protostar.
Our numerical modeling corroborates their conclusion -- Class 0/I disks develop 
vigorous gravitational instability that helps keep the disk mass well below that of the protostar.

To what extent can our numerical modeling overestimate disk masses?
The reason for the overestimate may be twofold.
First, Class II disk masses in Figures~\ref{fig2} and \ref{fig5} were
derived from disks which have ages of order 3~Myr but real disks may be older.
To test this possibility, we ran a few models for 5~Myr and compared the resulting 
Class II disk masses with those obtained for 3~Myr-old disks. We found that longer ages of disks 
could only cause a factor of 2 (at most) decrease in the time-averaged Class II disk masses,
still not enough to reconcile our obtained disk masses with those of AW.
Second, we might have not taken into account some important physical
mechanisms that help reduce disk masses. 
Indeed, dust disks and accretion disappear on time scales of about 6~Myr and
this points to the existence of additional disk clearing mechanisms, especially 
relative to the non-viscous disk models.
Magnetic braking is known to be efficient 
at transporting disk angular momentum directly to the external environment. 
However, disks are known to have low ionization fractions and 
ambipolar diffusion may strongly moderate the effect of magnetic braking.
We plan to explore this mechanism in a follow-up paper.
The formation of a binary system can also decrease the resulting disk masses
but the fraction of binary systems is under debate and it may be too low to reconcile our
model disk masses with observations \citep{Lada}. 
Photoevaporation of disks due to the external ultraviolet radiation may somewhat decrease 
disk masses but it operates only late in the evolution of Class II disks and is expected to
have little influence on the time-averaged disk masses.
Clearly,  the statistics on Class~II disks must depend on the 
existence of additional disk clearing mechanism and the adopted end time in our numerical and
more work is needed to resolve the cause of disparity between observationally and numerically
derived disk masses.

\section{Numerical and model caveats}
\label{caveats}
\subsection{Gas thermodynamics}
It has recently become evident that
the gas thermodynamics plays an important role in regulating gravitational instability 
and fragmentation in circumstellar disks \citep[e.g.][]{Pickett03,Johnson03,Rice03,Boley06,Cai08}. 
While most researches agree that circumstellar disks are susceptible to gravitational instability, 
particularly in the early phase of stellar evolution, 
there still no consensus on whether fragmentation
ever occurs and, if it does, whether any clumps that form will become bound protoplanets.
For instance, \citet{Johnson03} have considered 
geometrically thin disks that cool by radiatively transporting thermal energy (generated in the midplane)
to the disk surface. They have found that disks are susceptible to fragmentation only if 
the effective cooling time is comparable to or smaller than the dynamical time. Similar 
conclusions have been made by \citet{Rice03} and \citet{Mejia04} using global SPH and grid-based
numerical simulations, respectively. Fragmentation can be stabilized in the inner disks by slow cooling
\citep[e.g.][]{Rafikov, Stamatellos} and in the outer disks by stellar and envelope 
irradiation \citep{Matzner05,Cai08}.
On the outer hand, disk fragmentation can be aided by the infall of matter from an external envelope,
certainly in the early phase of disk evolution \citep{VB1,VB2,Krumholz,Kratter}. 

The accurate implementation of gas thermodynamics in numerical simulations
of self-consistent disk formation and long-term evolution is a very difficult task.
This is because our numerical simulations capture several physically distinct regimes 
that see a substantial change with time in the chemical composition, opacities, and dust properties.
Numerical techniques allowing for the accurate treatment of thermal physics 
in such numerical simulations are only starting to emerge \citep[e.g.][]{Stamatellos07b}.
Taking into account the complexity of gas thermodynamics in circumstellar disks, 
we use the barotropic equation of state.

 It should be stressed here that circumstellar disks described by the 
barotropic equation of state with $\gamma=1.4$ are susceptible to fragmentation and 
formation of stable clumps in the early embedded phase of evolution. These clumps are 
later driven onto the protostar and produce bursts of mass accretion that can help explain 
FU Ori eruptions. This phase of protostellar accretion is known as the burst phase, 
it operates in disk-star systems with mass $\ga 0.6~M_\odot$ 
and it is very efficient at transporting the disk material onto the protostar \citep{VB1,VB2}. 
Circumstellar disks described by $\gamma=5/3$ are hotter and less susceptible to
fragmentation but the clump production is not completely suppressed \citep{VB2}.
%A more realistic treatment of the thermal physics
%using radiation transfer tends to produce hotter disks and create less clumps 
%in the early embedded phase of disk evolution, though not completely terminating 
%their production. 
Recent numerical and semi-analytic modeling using a more accurate prescription for the energy
balance in the disk (including radiation transfer) also predict
clump formation in disks (particularly, in their outer parts) 
around stars with mass equal to or more massive than one solar mass
\citep{Krumholz,Mayer07,Stamatellos07,Boss08,Kratter}.

In order to examine if a higher disk temperature can affect the accretion history and, consequently,
disk masses, we stiffened the barotropic equation of state by rasing $\gamma$ from 1.4 to $5/3$.
In the case of self-gravitating disks, this affects most of the disk material
because most of the disk is optically thick and is characterized by surface densities 
considerably larger than $\Sigma_{\rm crit}=36.2$~g~cm$^{-2}$ \citep{VB3}. On the other hand, 
Class II viscous disks may have surface densities comparable to or lower than
$\Sigma_{\rm crit}$, particularly in the late evolution phase \citep{VB4}. Even in this case, the increase in $\gamma$ is expected to affect
a large portion of the disk material because equation~(\ref{barotropic}) 
makes a gradual transition between the optically thin $\Sigma << \Sigma_{\rm crit}$ and
optically thick $\Sigma>>\Sigma_{\rm crit}$ regimes. 
The detailed numerical simulations are presented in \citet{VB4}, here we provide only the main 
results. We find that while an increase in $\gamma$ raises the disk temperature 
from 50~K (at 10~AU) to about 100~K and  makes the disk less susceptible to fragmentation 
and clump formation, the amount of accreted material changes insignificantly. In fact,
the mass accretion rates time-averaged over $\sim 10^4$~yr are very similar in models with $\gamma=1.4$
and $\gamma=5/3$, while the instantaneous rates may be quite different.
This is because a modest increase in disk temperature acts to 
suppress higher order spiral modes $m > 2$, which are rather inefficient at transporting mass 
radially inward and tend to produce 
more fluctuations and cancellation in the net gravitational torque. At the same time,
the growth of lower order modes $m \le 2$ is promoted, 
which are efficient agents for mass and angular momentum redistribution. 
A similar effect was reported by \citet{Cai08} for circumstellar disks heated by a strong
envelope irradiation. Thus, the relative strength of lower order spiral modes increases 
in hotter disks, which compensates an apparent decrease
in the amount of accreted gas due to the less efficient burst phase of accretion. 

\subsection{Numerical resolution and central point mass}
% The radial points are logarithmically spaced.
%The innermost grid point is located at $r=5$~AU, and the size of the 
%first adjacent cell lies in the range between 0.26~AU (model~21) 
%and 0.36~AU (model~7). 
The use of a logarithmically spaced numerical grid in the radial direction means 
that the ratio $\Delta r/r$ of the cell size $\Delta r$ 
to radius $r$ is 
constant for a given cloud core and varies from $0.05$ (model~21) to 
$0.07$ (model~7). It can be noticed from Fig.~\ref{figA1} that the aspect ratio $A=z/r$
of the disk vertical scale height to radius is smaller than $\Delta r/r$ in the inner 70~AU, 
which implies that a better radial (and angular) resolution is desirable in this inner region. 
We have found that an increase in the resolution to $256 \times 256$ grid points makes 
little influence on the accretion history and disk masses but helps save a considerable amount 
of CPU time and, eventually, consider many more model cloud cores. We also point out that, due to
the adopted log spacing in the radial direction, the resolution in the inner computational
region is sufficient to fulfill the Truelove criterion in the disk.

In our numerical simulations the position of the central star is fixed in 
the coordinate center. In reality, however, the star may wobble around the center 
of mass in response to the gravity force of the disk. This
can promote the growth of non-axisymmetric spiral modes (especially the m=1 mode) 
in massive disks and, consequently, help limit disk masses in the early phase of stellar evolution
\citep{Adams89,Shu90}. We plan to explore this potentially important mechanism in 
a follow-up paper.

\subsection{The weight function}
The mean disk masses listed in Table~\ref{table4} may depend on the 
form of the weight function. Our adopted weight function 
${\cal F}_{\rm K}$ [see eq.~(\ref{function})] assumes that stellar masses in each
stellar evolution phase are distributed according to the Kroupa IMF \citep{Kroupa01},
which actually refers to the initial masses of stars that have already formed.
To determine the extent to which our mean disk masses depend on the particular form
of the weight function, we also consider 
weight functions with a slope typical for the Salpeter IMF, 
${\cal F}_{\rm S} \left( \langle M_\ast \rangle \right)=A\, 
\langle M_\ast\rangle^{-2.35}$ \citep{Salpeter}, and Miller-Scalo IMF \citep{Miller}
\begin{equation}
{\cal F}_{\rm MS}\left(\langle M_\ast \rangle\right) = \left\{ \begin{array}{lll} 
   A \, \langle M_\ast \rangle^{-1.25} & \,\,\, 
   \mbox{if $0.1~M_\odot < \langle M_\ast \rangle \le 1.0~M_\odot$ } \\ 
   B \, \langle M_\ast \rangle^{-2.0} & \,\,\,
   \mbox{if $1.0~M_\odot < \langle M_\ast \rangle \le 2.0~M_\odot$ } \\
   C \, \langle M_\ast \rangle^{-2.3} & \,\,\,
   \mbox{if $\langle M_\ast \rangle > 2.0~M_\odot$. } 
    \end{array} 
   \right. 
   \label{MillerScalo}  
 \end{equation}
The resulted mean disk masses are listed in Table~\ref{table5}. It is evident that
the use of different weight functions (indicated in parentheses) yields mean values that are 
different from those derived from the Kroupa law by $30\%$ at most.

\begin{table}
\begin{center}
\caption{Mean disk masses with different weight functions}
\label{table5}
%\vspace{3 pt}
\begin{tabular}{cccc}
\hline
\hline
disk type & $ \!\!\! \overline{M}_{\rm d,C0}$ & $\!\!\! \overline{M}_{\rm d,CI}$ & $\!\!\! \overline{M}_{\rm d,CII}$  \\ 
\hline
self-gravitating (${\cal F}_{\rm S}$) & 0.08 & 0.08 & 0.10  \\
viscous (${\cal F}_{\rm S}$) & 0.09 & 0.09 & 0.05  \\
self-gravitating (${\cal F}_{\rm MS}$) & 0.09 & 0.12 & 0.15 \\
viscous (${\cal F}_{\rm MS}$) & 0.10 & 0.13 & 0.08 \\
 \hline
\end{tabular} 
\tablecomments{All disk masses are in $M_\odot$. The weight functions used to derive
the mean disk masses are shown in parentheses.}
\end{center}
\end{table}

\section{Summary}
\label{summary}
We have considered numerically the long-term evolution of self-consistently formed 
self-gravitating and viscous circumstellar disks around low-mass stars ($0.2~M_\odot \la 
M_\ast\la 2.0~M_\odot$).
We seek to determine time-averaged disk masses in the Class~0 ($\langle M_{\rm
d, C0}\rangle$), Class~I ($\langle M_{\rm d,CI}\rangle$), 
and Class~II ($\langle M_{\rm d,CII} \rangle$) stellar evolution phases. 
Time averaging in each evolution phase is done over the duration of the phase.
The mean disk masses ($\overline{M}_{\rm d}$) are then derived from these time-averaged masses 
by weighting them over the corresponding stellar masses using a power-law function
with a slope typical for the Kroupa initial mass function of star (see eq.~[\ref{mean}]). 
In self-gravitating disks the radial transport of mass and angular momentum is performed 
exclusively by gravitational torques, while in viscous disks both the gravitational and viscous torques
are at work. Our numerical simulations yield the following results.

\begin{enumerate}
\item Both the self-gravitating and viscous disks have Class~I
mean masses $\overline{M}_{\rm d,CI}=0.10-0.11~M_\odot$ that are slightly more
massive than those of Class~0 disks $\overline{M}_{\rm d,C0}=0.09-0.10~M_\odot$.
However, viscous Class II disks have a mean mass ($\overline{M}_{\rm d,CII}=0.06~M_\odot$) that
is a factor of 2 lower than that of self-gravitating Class II disks.
This is explained by the fact that gravitational torques prevail through the Class 0 and Class I 
phases but succumb to viscous torques through much of the Class II phase.

\item Our obtained mean disk masses are {\it larger} than
those derived by AW for 153 YSOs in the Taurus-Auriga star formation region, 
regardless of the physical mechanisms of mass transport in the disk.
The difference is especially large for Class II disks, for which we find
$\overline{M}_{\rm d,CII}=0.06-0.12~M_\odot$ but AW report median masses 
of order $3\times 10^{-3}~M_\odot$.
Our Class~I disks have almost a factor of 4 larger mean disk mass than those of AW.

\item Time-averaged disk masses have a considerable scatter around mean values in each evolution phase.
For instance, Class~0 and Class~I disk masses lie in the range $\langle M_{\rm d,C0} \rangle = 0.04-0.2~M_\odot$
and $\langle M_{\rm d,CI} \rangle=0.02-0.55~M_\odot$, respectively, 
while Class II disks have a much wider range in
masses $M_{\rm d,CII}=0.004-0.7~M_\odot$.

\item When the three stellar evolution phases are considered {\it altogether}, 
we find a near-linear relation between time-averaged disk and stellar masses, 
$\langle M_{\rm d} \rangle = \left( 0.2 \pm 0.05 \right) \, \langle M_\ast \rangle^{1.3\pm 0.15}$.
This relation can potentially become somewhat shallower if cloud cores with very low
ratios of rotational-to-gravitational acceleration are considered, but cannot be broken completely.
However, when each phase is considered {\it separately}, the relation between disk and stellar 
masses becomes progressively steeper along the sequence of stellar evolution phases.
In particular, the corresponding relations for Class 0, Class I, and Class II objects have
exponents $0.7\pm 0.2$, $1.3\pm 0.15$, and $2.2 \pm 0.2$, respectively.

\item  In each stellar evolution phase, time-averaged disk-to-star mass ratios 
$\langle \xi \rangle$ tend to have greater values 
for stars of greater mass. However, there is a clear saturation effect -- $\langle \xi \rangle$ 
never exceed $40\%$, regardless of the stellar mass. 
{\it This means that no star can harbour a disk with mass comparable to or greater than
that of the star}.
This saturation effect is caused by the onset of vigorous
gravitational instability in circumstellar disks that form in the Class~0 or early Class~I phase
\citep[see also][]{Adams89,Shu90}.

\item Only low-mass objects $M_\ast \le 0.9~M_\odot$ are expected to have Class~0 disks.
Some Class~0 objects (particularly those formed from slowly rotating cloud cores)
may have no disks and outflows associated with them. 
%We predict that protostellar outflows do not form around these objects.

\end{enumerate}

\acknowledgments
   The author is thankful to the referee for suggestions and comments that 
   helped improve the manuscript and 
   Prof. Shantanu Basu for stimulating discussions. The author
   gratefully acknowledges present support 
   from an ACEnet Fellowship. Numerical simulations were done 
   on the Atlantic Computational  Excellence Network (ACEnet). Some of the simulations
   we done on the Shared Hierarchical Academic Research Computing Network (SHARCNET)
   while the author was a CITA National Fellow at the University of Western Ontario.

\appendix

\section{Disk scale height}
We derive the disk vertical scale height $Z$ at each computational cell 
via the equation of local vertical pressure balance 
\begin{equation}
\rho \, \tilde{c}_s^2 = 2\int_0^Z \rho \left( g_{z,\rm gas}+g_{z,\rm st} \right) dz,
\label{eq1}
\end{equation}
where $\rho$ is the gas volume density, $g_{z,\rm gas}$ and $g_{z,\rm st}$ are the {\rm vertical} 
gravitational accelerations due to self-gravity of a gas layer and gravitational pull of 
a central star, respectively. 
Assuming that $\rho$ is a slowly varying function of vertical distance $z$ between $z=0$ (midplane)
and $z=Z$ (i.e. $\Sigma=2\, Z \,\rho$) and using the Gauss theorem, one can show that
\begin{eqnarray}
\int_0^Z \rho \, g_{z,\rm gas} \,  dz &=& {\pi \over 4} G \Sigma^2 \label{eq2} \\
\int_0^Z \rho \, g_{z,\rm st} \, dz &=& {G M_\ast \rho \over r} 
\left\{  1-\left[ 1+ \left({\Sigma\over 2 \rho r} \right) \right]^{-1/2} \right\},
\label{eq3}
\end{eqnarray}
where $r$ is the radial distance and $M_\ast$ is the mass of the central star.
Substituting equations~(\ref{eq2}) and (\ref{eq3}) back into equation~(\ref{eq1}) we obtain
\begin{equation}
\rho \, \tilde{c}_s^2 = {\pi \over 2} G \Sigma^2 + {2 G M_\ast \rho \over r} 
\left\{  1-\left[ 1+ \left({\Sigma\over 2 \rho r} \right) \right]^{-1/2} \right\}.
\label{height1}
\end{equation}
This can be solved for $\rho$ given the model's known $\tilde{c}_s^2$, $\Sigma$, and $M_\ast$
using Newton-Raphson iterations.
The vertical scale height is finally derived as $Z=\Sigma/(2\rho)$.

Finally, we compare our model values of $Z$ with those  predicted 
from detailed vertical structure models of irradiated accretion disks around T Tauri stars 
by \citet{DAlessio}.  Their figure~1b (dashed curve) yields the following relation 
between the aspect ratio $A=Z/r$ of the local scale height to radial distance and  radial distance
$r$
\begin{equation}
A = 0.1 \left( r/100\,{\rm AU} \right)^{1/4},
\label{height2}
\end{equation}
which is plotted by the dashed line in Figure~\ref{figA1}.
The solid line in Figure~\ref{figA1} shows our obtained aspect ratio $A=Z/r$ for model~13 at t=0.3~Myr.
It is evident that our $A$ is somewhat underestimated in the
inner disk and overestimated in the outer disk but the disagreement between the two curves 
is within a factor of unity.

\begin{figure}
\centering
  \includegraphics[width=10 cm]{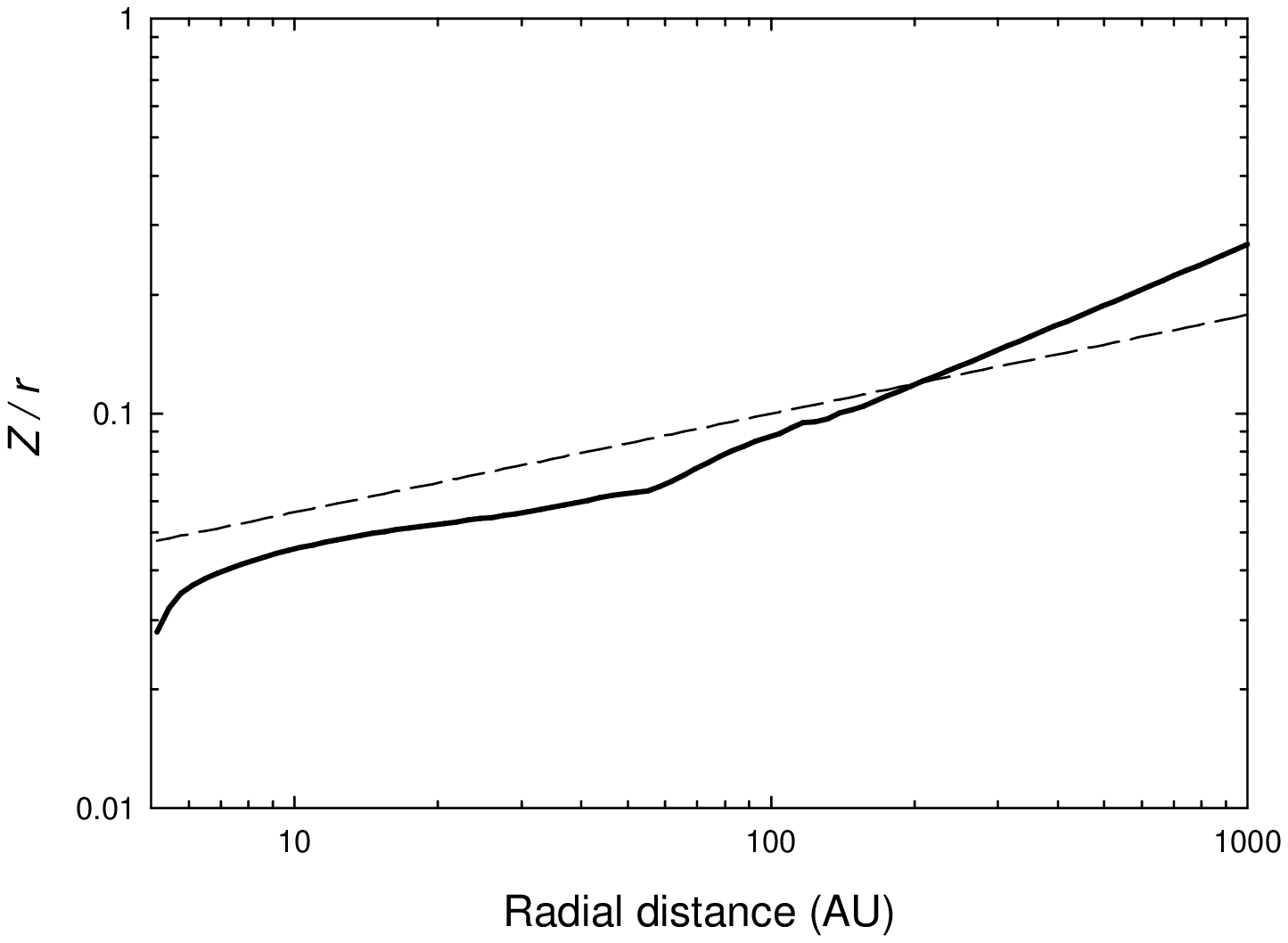}
      \caption{Aspect ratio of the disk vertical scale height to radius (Z/r)
      as a function of radius (r). The thick solid line shows Z/r
      calculated using equation~(\ref{height1}), while the dashed line does that
      for equation~(\ref{height2}).}
         \label{figA1}
\end{figure}

\section{Divergence of the viscous stress tensor}
The components of $\nabla \cdot {\bl \Pi}$ in polar coordinates ($r,\phi$) are
\begin{eqnarray}
\left( \nabla \cdot {\bl \Pi} \right)_r &=& {1\over r} {\partial \over \partial r} r \Pi_{rr} +
{1 \over r}  {\partial \over \partial \phi} \Pi_{r\phi} - {\Pi_{\phi\phi} \over r}, \\
\left( \nabla \cdot {\bl \Pi} \right)_\phi &=& {\partial \over \partial r} \Pi_{\phi r}
+ {1\over r} {\partial \over \partial \phi} \Pi_{\phi\phi} + 2 {\Pi_{r\phi} \over r},
\end{eqnarray}
where we have neglected the contribution from off-diagonal components $\Pi_{rz}$ and $\Pi_{\phi z}$.
The components of the viscous stress tensor $\bl \Pi$ in polar coordinates ($r,\phi$) can be found in
e.g. \citet{VT}.

\end{document}